\documentclass[12pt,aps,prb,preprint]{revtex4}

\usepackage{amsmath}
\usepackage{graphicx}
\begin{document}

\title{The Atwood's  machine as a tool to introduce variable mass systems}
\author{C\'elia A. de Sousa}
\affiliation{Departamento de F\'\i  sica da Universidade de Coimbra,\\ P-3004-516
Coimbra, Portugal}
%\email{celia@teor.fis.uc.pt}

\begin{abstract}

This paper discusses an instructional strategy which explores eventual  similarities
and/or analogies between  familiar problems and more sophisticated systems. In this
context, the Atwood's machine problem is used to introduce students  to  more complex
problems involving ropes and chains. The methodology proposed helps students to develop
the ability needed to apply relevant concepts in situations not previously encountered.
The pedagogical advantages are relevant for both secondary and high school students,
showing that, through adequate examples, the question of the validity of Newton's second
law may be  introduced to even  beginning students.

\end{abstract}
\vskip3cm

 \maketitle

\section{Introduction}

The description of the motion of a uniform rope  over a smooth pulley under the influence
of gravity can be used  to exemplify the behavior of systems with variable mass.
The discussion of variable mass  problems from the concept of momentum flux (e.g. Siegel
1972 and Sousa 2002) or from the generalization  of the Newton's second law (e.g. Sousa
and Rodrigues 2004) is not adequate for introductory physics.
However, the inter-disciplinarity   of this important subject and  the  relevance of its
  applications in rocket theory (e.g. Meirovitch 1970, Tran and Eke 2005), astronomy
(e.g. Kayuk and Denisenko 2004), biology (e.g. Canessa 2007, 2009), robotics (e.g.
Djerassi 1998), mechanical and electrical machinery (e.g. Cveticanin 2010), etc, justify
the introduction of this theme  to  students in the scientific areas as soon as possible.

We observe that sometimes  students have  relevant mathematical knowledge but fail to
apply or interpret that knowledge in the context of physics.
 This produces a barrier and an additional difficulty to the robust use of concepts in complex problem
 solving.
So, it is important to elaborate   strategies that can help students employ the
mathematical knowledge they already possess.

In a presentation to a group of pre-university students, we tried a new method to solve
problems of ropes or chains; it starts by   exploring   eventual similarities and/or
analogies with the most simple and familiar problem on the Atwood's machine.
The versatility of this system   has been confirmed by many generations of teachers
(Greenslade 1985). It provides a rich source of ideas for experiences and problems in the
application of Newton's second  law to the motion of a compound system, being analysed by
the generality of students of physics and engineering.

In the present paper, the Atwood's machine problem is solved by the traditional method
based on Newton's second law for particles (Method 1). After that, we verify the results
obtained by considering other methodologies, involving the concept of  centre of mass
(Method 2) and conservation of energy (Method 3), which are appropriate  to the more
complex problem of the rope.

%%%%%%%%%%%RRRR%%%%%%%%%%%%%%%%

\section{The Atwood's machine problem}

{\em Two  blocks of masses $m_1$ and $m_2$ ($m_2>m_1$) are connected by a  massless
string passing over a   frictionless   pulley of negligible mass. The mass $m_2$ is
released from rest at $t=0$ (figure 1. (a)). Find the acceleration of the blocks and the
tension in the string.}

\begin{figure}[h]
\begin{center}
{\includegraphics{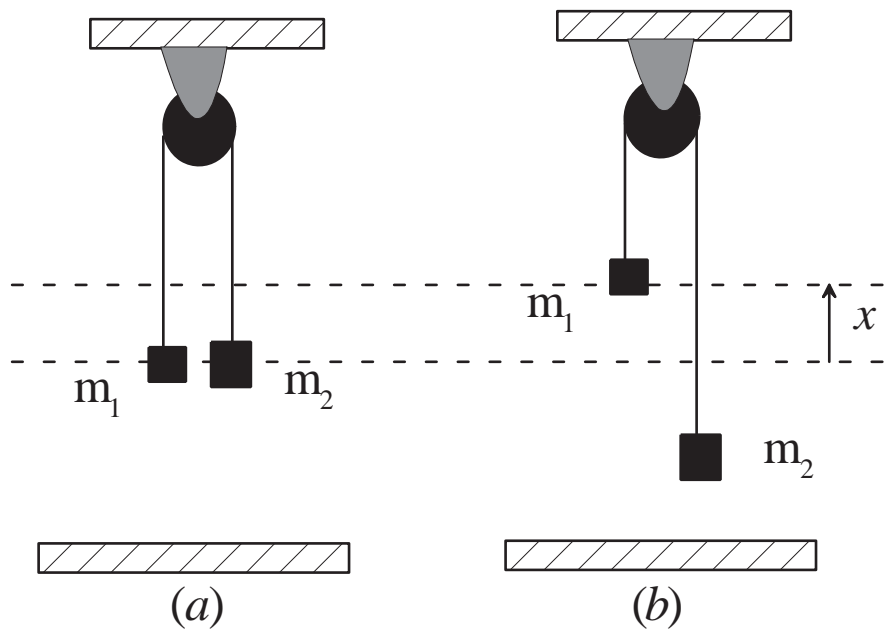}} \caption{ Atwood's machine.
  Configuration of the system at $t=0$ (a) and at $t\neq 0$ (b).}
\end{center}
\end{figure}

\vskip0.3cm

\noindent {\bf Method 1.}

 \vskip0.3cm

 The traditional analysis of this one-dimensional motion consists in the use of the
Newton's second law  for translation

\begin{equation}
\vec F\,=\,m\,\vec a,\label{newpart}
\end{equation}
where $\vec F$ is the sum of all forces acting on the particle with mass $m$ and
acceleration $\vec a$.

 The free-body diagram for each block contains the downward force of gravity and
the upward tension force $\vec T$ exerted by the string.  As the string is inextensible
both masses have  acceleration with equal magnitude  $a$ . As we assume $m_2>m_1$, the
object $m_1$ accelerates upward, and $m_2$ accelerates downward. As the motion of the
blocks is  one-dimensional in the vertical direction, there is no need to use vectors
explicitly.

Applying Newton's second law (\ref{newpart}) to blocks 1 and 2 we obtain, respectively,

\begin{equation}
T\,-\,m_1\,g\,=\,m_1\,a, \label{a2}
\end{equation}
and

\begin{equation}
m_2\,g\,-\,T\,=\,m_2\,a. \label{a1}
\end{equation}
From these two equations it is easy to find

\begin{equation}
a\,=\,g\,\frac{m_2\,-\,m_1}{m_1\,+\,m_2},\label{ac}
\end{equation}
and
\begin{equation}
T\,=\,2\,g\,\frac{m_1\,m_2}{m_1\,+\,m_2}.\label{T}
\end{equation}

The equations (\ref{T})  and  (\ref{ac}) satisfy  two special cases: when $m_1=m_2=m$,
$a=0$ and $T=m\,g$; if $m_2>>m_1$, then $a\simeq g$ and $T\simeq 2 \,m_1 \,g$.

At this point the teacher must remember that, in principle,  equation  (\ref{newpart}) is
not valid for variable mass systems.

\vskip0.3cm

\noindent {\bf Method 2.}

\vskip0.3cm

 The solution already obtained can be confirmed if we substitute
one of the equations (\ref{a2})  or  (\ref{a1})   by considering the system as being made
up of both objects. The tension in the string is now an internal force, and the external
forces acting on the system are the downward force of gravity and the upward force of
reaction by the pulley, $\vec N$, which magnitude is $N\,=\,2\,T$.

The Newton's second law for the translation of the centre of mass of the system of particles is given by:

\begin{equation}
\vec F\,=\,m\,\vec a_{\rm cm},\label{newsist}
\end{equation}
where $\vec F$ is the sum of all external forces acting on the total mass  of the system,
$m\,=\,m_1\,+\,m_2$, and $a_{\rm cm}$ is the acceleration of the centre of mass.

As the centre of mass moves downward we can write  the following equation of motion:

\begin{equation}
(m_1\,+\,m_2)\,g\,-\,2\,T\,=\,(m_1\,+\,m_2)\, a_{\rm cm},\label{newsyst}
\end{equation}
where the acceleration of the center of mass is given by

\begin{equation}
a_{\rm cm}\,=\,a\,\frac{m_2\,-\,m_1}{m_1\,+\,m_2}.\label{acm}
\end{equation}

The insertion of equation  (\ref{acm}) into (\ref{newsyst}), and using one of the
equations of motion  (\ref{a2})  or  (\ref{a1}),  the results already obtained by Method
1 are confirmed.

\vskip0.3cm

\noindent{\bf Method 3.}

\vskip0.3cm

Let us now analyse  the conservation of energy. Comparing the configuration of the system
at the  instant $t$ with those at $t=0$, we easily obtain an equation which determines
the velocity as a function of $x$ (see figure 1 (b)).

We consider that the potential energy, $U$, is zero in the configuration of the system at
$t=0$. Assuming that the blocks are initially at rest, the kinetic energy, $K$, is also
zero at $t=0$.  The friction is negligible and the conservation of energy states

\begin{equation}
K_0\,+\,U_0\,=\,K\,+\,U.\label{cen}
\end{equation}

As $K\,=\,\frac{1}{2}\,(m_1\,+\,m_2)\,v^2$ and $U\,=\,(m_2\,-\,m_1)\,g\,x$, by using
equation (\ref{cen}) the squared velocity at position $x$ of mass $m_1$ is found to be

\begin{equation}
v^2\,=\,2\,g\,x\,\frac{m_1\,-\,m_2}{m_1\,+\,m_2}.\label{v2}
\end{equation}

This equation allows to apply a mathematical procedure that students already know from
math classes, but that  do not usually apply in  classes of introductory physics. In
fact, this equation allows directly to  the acceleration $a$ by using the identity

\begin{equation}
a\,=\,\frac{1}{2}\,\frac{{\rm d}\,v^2}{{\rm d}\,x}.\label{id}
\end{equation}

This equation comes  from the following mathematical procedure:

\begin{equation}
a\,=\,\frac{{\rm d}\,v}{{\rm d}\,t}\,=\,\frac{{\rm d}\,v}{d\,x}\,\frac{{\rm d}\,x}{{\rm
d}\,t}\,=\, \frac{{\rm d}\,v}{{\rm d}\,x}\,v\,=\,\frac{1}{2}\,\frac{{\rm d}\,v^2}{{\rm
d}\,x}.
\end{equation}

Using equations  (\ref{v2}) and (\ref{id}) we easily recover the acceleration $a$, as it
must.

 A more common procedure to obtain the acceleration from equation (\ref{v2}) consists in using the
equations of motion for position and velocity as functions of time: $x\,=\,a\,t^2\,/2$
and $v\,=\,a\,t$. However, the  procedure here adopted  (less common for beginning
students) is more convenient for  cases where the acceleration is not constant as  in the rope  problem.

\section{The falling rope problem}

 {\em A uniform and flexible rope of length $l$, and mass per unit length $\lambda$ hangs almost symmetrically
over a frictionless and  small  pulley. Due to a small perturbation, the rope begins to
fall from rest at $x=0$. Find the velocity of the rope when it leaves the pulley, as well
as the acceleration of the rope. Figure 2 shows the configuration of the system at the
instant $t$.}

\vskip-0.2cm
\begin{figure}[h]
\begin{center}
{\includegraphics{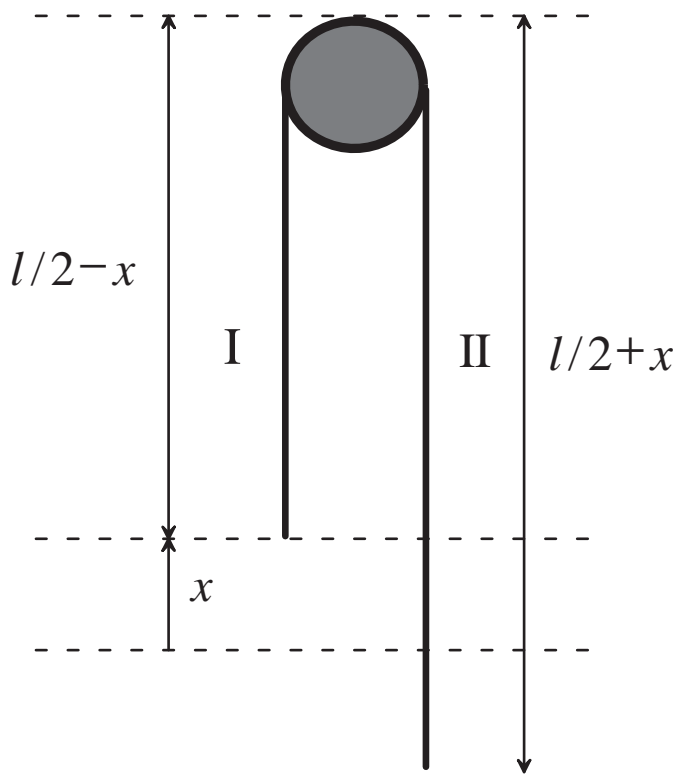}} \caption{ Configuration of the system at $t\neq 0$.}
\end{center}
\end{figure}
\vskip-0.2cm

  In analogy with the Atwood's machine problem, we can divide the whole system
(constant mass) in two sub-systems I and II (variable mass). If we want to keep the
problem suitable for beginning students, we should not   consider
 Method 1 based on Newton's second law for  sub-systems I and II.
However, some aspects of this subject can be discussed {\em a posteriori}, following the
methodology presented in the Appendix.

The radius of the pulley, quite big  in the figure, is supposed to be very small compared
with the length $l$ of the chain. This  means that the movement is, in  good
approximation, one-dimensional in the vertical direction, and, also in this problem,
there is no need to use vectors explicitly.

Let us start with the conservation of energy.  To this purpose we define the position of
the centre of mass  by using the  expression

\begin{equation}
x_{\rm cm}\,=\,\frac{x_I\,m_I\,+\,x_{II}\,m_{II}}{\lambda\,l},\label{xcm}
\end{equation}
where $m_I\,=\,\lambda\,(l/2\,-\,x)$, $m_{II}\,=\,\lambda\,(l/2\,+\,x)$,
$x_I\,=\,l/4\,+\,x/2$ and $x_{II}\,=\,l/4\,-\,x/2$. The variable $x$ denotes the
displacement of one end of the rope from its initial position as indicated in figure 2
($0<x<l/2$).

The previous equations allow to obtain
\begin{equation}
x_{\rm cm}\,=\,\frac{l}{4}\,-\,\frac{x^2}{l}.\label{xcm}
\end{equation}

The equation (\ref{xcm}) satisfy  two special cases: when $x=0$, $x_{\rm cm}\,=\,l/4$; if
$x=l/2$, then $x_{\rm cm}\,=\,0$.

The mechanical  energy  initially $(x=0)$ and at a generic configuration of the system
($x\neq 0$) are given by

\begin{equation}
K_0\,+\,U_0\,=\,\lambda\,g\,\frac{l^2}{4},\label{em0}
\end{equation} and
\begin{equation}
K\,+\,U\,=\,\frac{1}{2}\,\lambda\,l\,v^2\,+\,\lambda\,g\,\left(\frac{l^2}{4}\,-\,x^2
\right).\label{emt}
\end{equation}

The velocity as a function of $x$ follows explicitly from the conservation of energy (\ref{cen})

\begin{equation}
v\,=\,x\,\left(\frac{2\,g}{l}\right)^{1/2}.\label{vro}
\end{equation}

Combining  the expression of $v^2$ with equation (\ref{id}) yields to the acceleration
\begin{equation}
a\,=\,2\,g\,\frac{x}{l}.\label{acx}
\end{equation}

In order to test the validity of the expression (\ref{vro}) in the interval $0<x<l/2$, we
must calculate the force of reaction by the pulley, $\vec N$. To this purpose we consider
Method 2 of the Atwood's machine problem.

The velocity and the acceleration of the centre  of mass, which moves downward, must be
calculated. We obtain successively,

\begin{equation}
v_{\rm cm}\,=\,\frac{v\,m_{II}\,-\,v\,m_{I}}{\lambda\,l}\,=\,2\,v\,\frac{x}{l},
\end{equation}
and

\begin{equation}
a_{\rm cm}\,=\,\frac{{\rm d}\,v_{\rm cm}}{{\rm
d}\,t}\,=\,2\,a\,\frac{x}{l}\,+\,2\,\frac{v^2}{l}=\,8\,g\, \frac{x^2}{l^2},\label{acmro}
\end{equation}
where equations (\ref{vro})and (\ref{acx}) have also been used to obtain the last equation.

The Newton's second law for the system of particles (\ref{newsist}),  applied to the
whole rope, allows to write

\begin{equation}
\lambda\,l\,g\,-\,N\,=\,\lambda\,l\,a_{\rm cm}\,=\,8\,\lambda\,g\,\frac{x^2}{l}.
\end{equation}

Therefore we may obtain the normal force

\begin{equation}
N\,=\,N\,(x)\,=\,\lambda\,l\,g\,\left(1\,-\,8\,\frac{x^2}{l^2} \right),\label{normal}
\end{equation}
allowing to conclude that the solution given by equation (\ref{vro}) is valid only till
the value $x\,=\,l/(2\,\sqrt{2})$, where the normal force attains the value $N=0$ (Calkin
1989). This fact  can be demonstrated   to students in the classroom, by observing some
whip-lashing behavior of the rope before its whole length  goes over the pulley.

It is interesting to notice that the expression for the acceleration (\ref{acx}),
although dependent on the coordinate $x$, is consistent  with the expression for the
acceleration in the Atwood's machine (\ref{ac}), i. e., satisfies to the relation
$a\,=\,g\,(m_{II}-m_{I})/(m_I+m_{II})$ in the referred interval of $x$. However, as the
mass of the rope is different from zero, an analogous equation for the tension in the
string is not verified, i.e., $T\neq 2\,g\,m_I\,m_{II}/(m_I\,+\,m_{II})$ (see the
Appendix).

So, the teacher can  also guide students to focus on important differences between both
problems. We point out:
\begin{itemize}
  \item Sub-systems I and II of variable mass can be described by the equation of motion $F=m\,a$, but $F=dp/dt$
  is not satisfied (see the Appendix).  The blocks of the Atwood's machine can be described by both forms of the
Newton's  second law.
  \item The forces acting on the small piece of the rope over the pulley satisfies  the
  condition $2\,T\,-\,N\,=\,0$, in the Atwood's machine problem, whereas $2\,T\,-\,N\,\neq 0$, in
  the rope problem (see the Appendix). This fact also gives the opportunity to enlarge the discussion to the concept
of  tension when the mass of the rope is nonzero.
\end{itemize}

In our opinion, the teacher must leave the complete analysis of the problem, from the  point of view
of a variable mass system, for  students in intermediate courses of mechanics.

\section{Conclusions}

The Atwood's machine problem solved by the traditional methodology does not require a
sophisticated level of mathematics. However, the enlargement suggested in the present
paper indicates that students' understanding  can be probed more deeply. In this way they
will be more prepared to re-create  the physical situation under study, or  make  a model
to better visualize the problem, or even think on analogous problems they have solved
before.
This strategy can guide the design of instruction to match the needs and performances of
students in introductory physics.

 In conclusion, the  present strategy provides an elegant way to solving certain aspects of the rope problem.
 However, at the same time, leaves aside other equally important aspects which may be  discussed in intermediate
 courses of mechanics.
Finally, we remark that similar problems, such as a massive rope sliding freely over a smooth nail, can be
 treated along the lines indicated in the present work (Sousa and Rodrigues 2004).
 Another example would be the motion of a chain, part of which is hanging off the edge of
 a smooth table.

{\bf Acknowledgements}

This work was supported by FCT.

%%%%%%%%%%%%%%%%%%%%%%%%%%%%%%%%%%%%%%
 {\bf Appendix}

 This Appendix applies the Newton's second law for variable mass systems. As the motion is one-dimensional, this
 equation of motion can be written in the form (Sousa and Rodrigues 2004)

\begin{equation}\label{eq:var}
\frac{{\rm d}\,{ p}}{{\rm d}\,t}\,=\,{ F}\,+\,{ u}\,\frac{{\rm d}\,m}{{\rm d}\,t},
\end{equation}
where  $m$ is the  {\em instantaneous} mass  and  ${ p}\,=\,m\,{v}$
 its linear momentum,
  ${ F}$   is the net external force acting upon the {\em variable} mass system, and
 ${ u}\,{{\rm d}m}/{{\rm d} t}$  is the rate at which momentum is
carried into or away from the system of mass $m$.

As in this case the velocity of the mass being transferred between the sub-systems is
$u=v$, and both sub-systems have  the  velocity $v$, equation (\ref{eq:var}) can be
re-arranged, confirming that a general equation of  type $F=m\,a$ applies to both
sub-systems.

Denoting by  $T$  the tension in the rope at the pulley, the equations of motion of
sub-systems I and II read,

\begin{equation}\label{II}
T\,-\,\lambda\,g\,( \frac{l}{2}\,-\,x)\,=\,\lambda\,g\,( \frac{l}{2}\,-\,x)\,a,
\end{equation}
and
\begin{equation}\label{I}
\lambda\,g\,( \frac{l}{2}\,+\,x)\,-\,T\,=\,\lambda\,g\,( \frac{l}{2}\,+\,x)\,a.
\end{equation}
Using the expression of the acceleration (\ref{acx})  we easily obtain
 the tension

\begin{equation}\label{Ten}
T\,=\,\frac{1}{2}\,\lambda\,g\,l\,(1\,-\,4\,\frac{x^2}{l^2}).
\end{equation}

The resultant force on the small piece of rope over the pulley $2\,T\,-\,N$ can then be
calculated using (\ref{normal}) together with the last equation, giving

\begin{equation}\label{piece}
2\,T\,-\,N\,=\,4\,\lambda\,\,g\,\frac{x^2}{l}\,=\,2\,\lambda\,v^2.
\end{equation}

This result shows that in the interval $\triangle t$ the momentum of this massive small
piece of rope ($\lambda\,\triangle x$) increases by $2\,\lambda\,v^2\,\triangle t$
downward.

%%%%%%%%%%%%%%%%%%%%%%%%%%%%%%%
{\bf References}

Siegel S 1972 More about variable mass systems  {\em  Physics Teacher} {\bf 40} 183

Sousa C A 2002 Nonrigid systems: mechanical and thermodynamic aspects {\em Eur. J. Phys.}
{\bf 23} 433

Sousa C A and Rodrigues V H  2004 Mass redistribution in variable mass systems {\em Eur.
J. Phys.} {\bf 25} 41

Meirovitch L 1970 General motion of a variable-mass flexible rocket with internal flow
{\em J. Spacecr. Rockets} {\bf 7} 186

Tran T and Eke F O 2005 Effects of internal mass flux on the attitude dynamics of
variable mass systems {\em Adv. Astronaut. Sci.} {\bf 119} (Issue Suppl.) 1297

Kayuk  Ya F and Denisenko V I 2004  Motion of a mechanical system with variable
mass-inertia characteristics {\em Int. Appl. Mech.} {\bf 40} 814

Canessa E 2007 Modeling of body mass index by Newtons's second law {\em Jour. Theor.
Biology} {\bf 248} 646

Canessa E 2009 Stock market and motion of a variable mass spring  {\em Physica A} {\bf
388} 2168

Djerassi S 1998 An algorithm for simulation of motions of variable mass systems {\em Adv.
Astronaut. Sci.} {\bf 99}  461

Cveticanin L 2010  Dynamics of the non-ideal mechanical systems: A review {\em J. Serbian
Soc. for Comp. Mech.} {\bf  4}  75

 Greenslade Jr T B 1985 Atwood's machine {\em  Physics Teacher} {\bf 23} 24

 Calkin M G
1989 The dynamics of a falling chain: II {\em Am. J. Phys.} {\bf 57} 157

\end{document}